\documentclass[preprint,overfull]{epl}
\newcommand{\e}  { {\rm e}}
\newcommand{\lb}  { l_{\rm B}}
\newcommand{\dd}  { {\rm d}}

\usepackage{amssymb}

\title{Polyelectrolyte Persistence Length: Attractive
Effect of Counterion Correlations and Fluctuations}
\shorttitle{Polyelectrolyte Persistence Length}
\author{Gil Ariel \and David Andelman}
\shortauthor{Gil Ariel \and David Andelman}
\institute{School of Physics and Astronomy \\
Raymond and Beverly Sackler Faculty of Exact Sciences \\
Tel Aviv University, Tel Aviv 69978, Israel}
\date{7/30/2002}

\begin{document}

\maketitle

\begin{abstract}

The persistence length of a single, strongly charged, stiff
polyelectrolyte chain is investigated theoretically. Path
integral formulation is used to obtain the effective
electrostatic interaction between the monomers. We find
significant deviations from  the classical Odijk, Skolnick and
Fixman (OSF) result. An induced attraction between monomers is
due to thermal fluctuations and correlations between bound
counterions. The electrostatic persistence length is found to be
smaller than the OSF value and indicates a possible mechanical
instability (collapse) for highly charged polyelectrolytes with
multivalent counterions. In addition, we calculate the amount of
condensed counterions on a slightly bent polyelectrolyte. More
counterions are found to be adsorbed  as compared to the Manning
condensation on a cylinder.
 \end{abstract}

Polyelectrolytes (PEs) are polymers that have ionizable groups.
When dissolved in solution, they dissociate into charged polymer
chains and a cloud of free, mobile counterions carrying opposite
charges \cite{BarratandJoannyreview,Oosawa}. Such macromolecules
appear in numerous industrial  applications as well as in
biological systems, introducing a new kind of biologically
inspired electrostatics \cite{physics_today}. The spatial
conformation of a single PE chain has been at the focus of
attention of experiments
\cite{experiments_list,BloomfieldDNAcondensation}, simulations
\cite{simulations_list,StevensKremerSimulation} and theoretical
models
\cite{models_list,Odijk,SkolnickFixman,FixmanLeBret,Shklovskii,Golestanian},
showing a wide range of behavior. The delicate balance between
counterion entropy and long range electrostatic repulsion can have
opposite effects on different PEs. Broadly speaking, in systems
containing weakly charged PEs and monovalent counterions, the
electrostatic repulsion dominates and makes the chains stiffer
\cite{FixmanLeBret,OSFworks}. On the other hand, highly charged
PEs with multivalent counterions experience an effective
attraction \cite{collapse} which is not well understood. This
attraction leads to enhanced flexibility and, in some cases,
induces collapse into a globular conformation \cite{collapse}. For
DNA macromolecules, this is known as DNA condensation
\cite{BloomfieldDNAcondensation}.

On a mean-field level, the Poisson-Boltzmann (PB) theory predicts
only intra-chain repulsion \cite{FixmanLeBret}. Another important
prediction is the Manning condensation \cite{Oosawa,Manning},
where some of the counterions are loosely bound to the PE chain.
Models going beyond mean-field theory take into account
correlations and thermal fluctuations
\cite{Shklovskii,Golestanian}. Correlations between bound
counterions become more significant at lower temperatures, where
the ions are considered to be arranged in a periodic fashion
similar to a Wigner crystal of electronic systems (or a
correlated liquid) \cite{Shklovskii}. At high temperatures, such
correlations are smeared out and become less important, while
counterion thermal fluctuations get larger and induce
an attraction (similar to van der Waals interactions),
which competes with the
usual repulsion between like charges
\cite{Golestanian}.

Despite an ongoing discussion, there is still no consensus on
which of the above mechanisms is more significant in physical
situations as realized in experiments
\cite{Corelations_vs_fluctuations}. In this letter, we propose a
field theoretical approach, which takes into account {\it both}
correlations and thermal fluctuations. This allows a consistent
examination of the two contributions at intermediate temperatures
and different charge densities of the polymer.

In order to account for counterion condensation we employ a
two-phase model \cite{Oosawa}: free counterions in solution are in
equilibrium with a one-dimensional gas of counterions bound to the
polymer backbone. In systems without added salt, the parameter
regulating counterion condensation is $q \equiv z \lb/a$ where
$z$ is the counterion valency, and $e/a$ is the monomer  linear
charge density for monomers of size $a$ and  charge $e$. The
Bjerrum length is defined as $\lb=e^2/\varepsilon k_B T$, where
$\varepsilon$ is the dielectric constant and $k_B T$ is the
thermal energy. We will see that $q$ is the significant
(temperature dependent) parameter which determines the system
behavior. On an infinite and straight cylinder Manning
condensation occurs for $q\ge 1$, and the condensed ions lower the
average charge density on the cylinder to an effective $q_{\rm
eff}=1$ \cite{Manning}. Below we will show how the condensation
picture changes for a semi-flexible PE modeled by a bent cylinder.

We model the polymer as a semi-flexible, worm-like chain of $N$
monomers. The persistence length, $l_{\rm p} \gg a$, is a measure
of the chain flexibility and only the case of a stiff, rod-like
polymer, $l_{\rm p} \gg L \equiv N a$, is considered. It is
generally agreed that once condensation is taken into account,
the overall charge density on the PE is small, and the effect of
free ions is to screen electrostatic interactions
\cite{StevensKremerSimulation,ZimmLeBret}. The inverse
Debye-H\"uckel screening length is defined as $\kappa^{-1}=  {[4
\pi z(z+1) \lb c]}^{-1/2}$, where $c$ is the concentration of
$z:1$ salt. The screening length is assumed to be much smaller
than the polymer contour length, $\kappa^{-1} \ll N a$ (as is
usually the case in experiments), and much larger than the
monomer size, $\kappa^{-1} \gg a$ (ensuring the validity of the
continuum approach employed here). We denote the spatial
conformation of the polymer as ${\bf R}(s), 0 \leq s \leq L$, and
the positions along the chain of the  $I$ bound counterions   as
${\bf R}(s_1) \dots {\bf R}(s_I)$. Up to a normalization factor,
the grand-canonical partition function of the system is
\begin{equation}
 {\mathcal Z} = \int  {\mathcal D} {\bf R}(s) \left( \sum_{I=0}^{\infty}
    \frac{ \e^{\mu I} }{I !}  \prod_{i=1}^{I}
    \frac{1}{L} \int_0^L \dd s_{i}     \right)
\e^{-H_0 - H_{\rm int}},
\label{partition_function}
\end{equation}
where the path integral is a sum over all possible spatial
conformations of the chain, $\mu$ is the chemical potential of a
one-dimensional gas of bound counterions, $H_0$ is the
Hamiltonian of a neutral chain with bare persistence length
$l_0$, and $H_{\rm int}$ is the Hamiltonian of the screened
electrostatic interaction between all charged monomers and bound
counterions. The monomer charges are assumed to have a uniform
charge density $e/a$ along the chain, while the counterions are
taken as point-like charges
\begin{eqnarray}
   H_{\rm int} &=& \frac{1}{2}
   \int_0^L ds \int_0^L \dd s^{\prime}
   U ({\bf R}(s)-{\bf R}(s^{\prime} ) )
\nonumber \\
     &\times &\Bigl[ z \sum_{i=1}^I \delta (s-s_i) - \frac{1}{a} \Bigr]
                      \Bigl[  z\sum_{j=1}^I \delta (s^\prime-s_j) - \frac{1}{a} \Bigr],
\label{H_int}
\end{eqnarray}
where $U({\bf r})= \lb \e^{- \kappa r}/r$ is the screened
electrostatic interaction in units of $k_{\rm B} T$. Because
$H_{\rm int}$ of Eq.~(\ref{H_int}) contains also
self-interactions, all following integrations   have a lower
cut-off at a distance of order $a$ . Continuous concentrations
are introduced in the following way
\begin{eqnarray}
    \phi^{\rm m} ( {\bf r} ) &=& \frac{1}{a} \int_0^L \dd s\, \delta (
                   {\bf r} - {\bf R} (s) )
\nonumber \\
    \phi^{\rm b} ( {\bf r} ) &=& \sum_{i=1}^{I}  \delta ( {\bf r} - {\bf R} (s_i) ).
\label{phi_def}
\end{eqnarray}
These can be substituted into the partition function
Eq.~(\ref{partition_function}) by making use of the identity
operator which couples the discrete and continuous
concentrations.  This is done using the path integral
representation of the delta function
\cite{Netz_review,auxiliary_fields}. The extra complexity of this
method is the introduction of two new auxiliary fields, denoted
$\psi^{\rm m}$ and $\psi^{\rm b}$ which couple to $\phi^{\rm m}$
and $\phi^{\rm b}$, respectively. The partition function then
reads
\begin{eqnarray}
      {\mathcal Z}  &=&  \int {\mathcal D} {\bf R} (s) \left( \prod_{i={\rm m,b}}
      {\mathcal D} \phi^i {\mathcal D} \psi^i \xi_i [{\bf R}] \right) \exp (-H_{\rm cont}),
\nonumber \\
         \xi_{\rm m} [ {\bf R} ] &=& \exp \left[ -H_{\rm id} + \right.
               \frac{i}{a} \int_0^L \dd s
            \left. \psi^m ( {\bf R} (s)) ~ \right]
\nonumber \\
         \xi_{\rm b} [ {\bf R} ] &=& \exp \left\{
           \int \dd^3 {\bf r} \left[  \frac{ \e^{\mu} }{N} \e^{  i \psi^{\rm b} (  {\bf r} ) }
           \phi^{\rm m} ( {\bf r} )
      + i {\bf \Phi} \cdot {\bf \Psi}
      \right] \right\}
\nonumber \\
     H_{\rm cont} &=&  \frac{1}{2} \int \int \dd^3 {\bf r}\, \dd^3 {\bf
     r}^{\prime}\,
        {\bf \Phi} ({\bf r}) \hat{Z}
          {\bf \Phi} ({\bf r^{\prime}})  U( {\bf r}- {\bf r}^\prime )
\nonumber \\
     {\bf \Phi} &=& { \phi^{\rm m} \choose \phi^{\rm b}  }  ,~
      {\bf \Psi}={ \psi^{\rm m} \choose \psi^{\rm b}  },~
        \hat{Z}  =
        \left(    \begin{array} {*{3}{c@{\: \:}}c@{\; \;}c}
           1  & -z  \\
           -z  & z^2
                  \end{array}   \right)
\label{cont_z}
\end{eqnarray}
where the vectors ${\bf \Phi}$ and ${\bf \Psi}$ and the matrix
$\hat{Z}$ are introduced to simplify notations. Next, $\xi_{\rm
b}$ is expanded in powers of $\psi^{\rm b}$ and the integrations
over $\phi^{\rm b}$ and $\psi^{\rm b}$ can be performed. The
effective electrostatic interaction $H_{\rm eff}$ and average
values of the different fields can be obtained by comparing the
partition function of Eq.~(\ref{cont_z}) with that of a neutral
system ($e=0$)  \cite{future_paper}. Thus, there is no need to
integrate out the polymer degrees of freedom $\{\phi^{\rm m}\}$
and $\{\psi^{\rm m}\}$. The method is similar to loop expansion in
field theory \cite{Netz_review}. The chemical potential $\mu$ is
set so that at the {\it straight rod} conformation, the average
number of bound counterions per unit length is $n^{\rm M}=(q-1)/
z^2 \lb=[a^{-1}-{(z \lb)}^{-1}]/z$, as predicted by Manning
\cite{Manning}. Since the bound counterion phase is in
equilibrium with free counterions in solution, $\mu$ does not
depend on the conformation of the polymer. In our model, as the
PE bends, the number of bound counterions is adjusted accordingly
in order to maintain this equilibrium.

Fixing the chemical potential of the bound counterion gas, rather
than the average density, has greater resemblance to experimental
systems, since the ions are only loosely bound and not chemically
attached to the polymer. Expanding Eq.~(\ref{cont_z}) to first
order in $\psi^{\rm b}$ (this is in fact a Gaussian approximation
\cite{future_paper}), $\mu$ turns out to be the chemical potential
of an {\it ideal} one-dimensional gas, $\mu=\ln (n^{\rm M} L)$,
and the total charge density does not depend on the conformation
of the polymer. The effective interaction Hamiltonian, $H_{\rm
eff,1}$ is found to be simply the screened electrostatic interaction between
all charges, which are uniformly smeared along the polymer at a
constant density, $1/z\lb$.
\begin{equation}
    H_{\rm eff,1} = \frac{1}{ 2 z^2 \lb }
       \int_0^L \dd s \int_0^L \dd s^\prime\,
       \frac{ \e^{ -\kappa | {\bf R}(s) - {\bf R}(s^\prime) | } }{ | {\bf R}(s) - {\bf R}(s^\prime) | }.
\label{heff1}
\end{equation}
This is just the Hamiltonian assumed by Odijk, Skolnick and Fixman (OSF)
\cite{Odijk,SkolnickFixman}, and their expression for the
persistence length can be
easily reproduced
\begin{equation}
    l_{\rm p} = l_0 + l_{\rm OSF} = l_0 + \frac{1}{4 z^2 \kappa^2 \lb},
\label{OSF}
\end{equation}
where $l_0$ is the bare persistence length of the neutral polymer
backbone and $l_{\rm OSF}$ is the electrostatic contribution.

Corrections to this approximation are obtained
through a cumulant expansion in higher powers of $\psi^{\rm b}$.
The second order expansion provides a 3-body
correction to the effective
interaction Hamiltonian \cite{future_paper}.
\begin{eqnarray}
   && H_{\rm eff,2} =
        -  { \e^\mu } \frac{z^2 \lb^2 }{2 L^3 } ( N-z \e^\mu )^2
\nonumber \\ &&
       \int_0^L\int_0^L\int_0^L\dd s\,\dd s^\prime\,\dd s^{\prime \prime}
       \frac{ \e^{ -\kappa | {\bf R}(s) - {\bf R}(s^\prime) | } }{| {\bf R}(s) - {\bf R}(s^\prime) | }
       \frac{ \e^{ -\kappa | {\bf R}(s) - {\bf R}(s^{\prime \prime}) | } }
              { | {\bf R}(s) - {\bf R}(s^{\prime \prime}) | }
\label{heff2}
\end{eqnarray}
where $\mu$ deviates from its ideal gas value and satisfies
\begin{eqnarray}
    && L n^{\rm M} \e^{-\mu} =
\nonumber \\ &&
1 + q ( 1-z\e^\mu /N )
    \int_0^L \dd s \int_0^L \dd s^\prime
    \frac{ \e^{ -\kappa | {\bf R}(s) - {\bf R}(s^\prime) | } }{ | {\bf R}(s) - {\bf R}(s^\prime) | } .
\end{eqnarray}
Note that the Manning condensation  for an infinite straight
cylinder is not expected to change substantially beyond its
mean-field value as was discussed in Ref.\cite{HaLiu}. Following
the method used by Odijk \cite{Odijk}, the electrostatic
persistence length $l_\e$ can now be calculated using the new
Hamiltonian, $H_{\rm eff,1}+H_{\rm eff,2}$. For low salt
concentrations ($\kappa a\ll 1$), the persistence length can be
expanded in powers of $1/\ln \kappa a$ yielding
\begin{equation}
    l_\e= l_{\rm OSF}
      \left[ q (2-q) - \frac{ (q-1)^2 }{q \ln (\kappa a)}
      + {\rm O} \left( 1 /[\ln (\kappa a)]^2 \right)  \right].
\label{el_pl}
\end{equation}
Equation~(\ref{el_pl}) is our main prediction and is depicted
in Fig.~1 for different counterion valencies $z=1,2,3$ as a
function of $\kappa a$. At low salt concentrations ($\kappa a\ll
1$) or high $q$,  the persistence length maintains the OSF
$\kappa^{-2}$ dependence, $l_\e\sim l_{\rm OSF}\sim \kappa^{-2}$.
We find that the electrostatic persistence length depends strongly on
the valency of the counterions.
For monovalent counterions, $l_\e$ is usually positive, indicating
an effective repulsion between the monomers.
However, its value is smaller than the one predicted by OSF.
Introduction of multivalent counterions reduces the rigidity of
the PE significantly and usually $l_\e<0$, indicating an effective
attraction between monomers.

Although the model assumes a rod-like PE, we  speculate that a PE
mechanical instability can be associated with $l_{\rm
p}=l_0+l_\e=0$, which represents the limit of validity of
rod-like behavior. In Fig.~1 it can be seen that with trivalent
counterions, PEs with a wide range of bare persistence lengths,
$l_0$, will be in a collapsed, globule-like conformation ($l_{\rm
p}<0$). However, we note that a detailed analysis of the chain
mechanical instability requires  methods different from the
persistence length prescription used in this Letter, to correctly
account for the collapse transition.

Using the same method, we performed an expansion to higher orders
of $\psi^{\rm b}$. Equation~(\ref{el_pl}), valid to second order,
accounts for most of the deviation from the OSF result. Third and
fourth order terms in the expansion represent only a relatively small correction to
the second-order.
This point will be discussed in greater
detail in a future publication \cite{future_paper}.
The expansion procedure  employed here is
asymptotic. As such, the approximation improves as  higher order
terms are taken into account up to a certain order. Beyond this
order the expansion diverges.

The leading difference between our $l_\e$ and $l_{\rm OSF}$ is
through a prefactor which depends on $q(=z\lb/a)$, and which  may
become negative at large $q$. The boundary between repulsive
(positive) and attractive (negative) electrostatic contributions
to the persistence length occurs at $q=2$, although it is not a
true phase transition since the total persistence length $l_{\rm
p}=l_0+l_\e$ is still positive. Taking into account higher order
corrections of Eq.~(\ref{el_pl}) may result in a shifted
threshold of $q=2$.

In addition, it is of interest to note that $q=2$ is the boundary
between fluctuation- and correlation-dominated regimes. The
length at which the average interaction between bound counterions
is equal to $k_{\rm B} T$ is $z^2 \lb$. This concentration of
bound counterions corresponds to $q=2$. It is expected that for
low values of the bound counterion concentration, $1<q<2$,  where
the average electrostatic interaction between ions is smaller
than $k_{\rm B} T$, thermal fluctuations will dominate over the
electrostatic, zero temperature ordering.  For higher
concentrations, $q \gg 2$,  models relying on the Wigner crystal
picture should be more appropriate. The boundary between the two
regimes is at $q=2$. As attraction is associated with
correlations, this comparison gives a new meaning to the $q=2$
threshold. The two complementary observations show that
fluctuations by themselves are not enough to induce effective
attraction. This argument, by no means, is a rigorous proof as it
does not capture precise numerical prefactors.

It is now possible to compare the persistence length of
Eq.~(\ref{el_pl}) with two previous models:  one which takes into
account thermal fluctuations \cite{Golestanian} and another for
correlations (correlated liquid) \cite{Shklovskii}. The
comparison with the fluctuation model is rather straightforward.
It consists of rewriting Eq.~(7) of Ref. \cite{Golestanian} in
terms of $l_{\rm OSF}$ and $q$ and expanding it in the two limits
$q\gg 1$ and $q \gtrsim 1$. In order to compare with the
correlation model (at finite $\kappa$), we slightly modified the
method used by Nguyen et al \cite{Shklovskii} to account for
finite salt concentration. Equation (13) of Ref.
\cite{Shklovskii}  describes the interaction of a counterion with
a single Wigner-Seitz cell in the absence of added salt. We
extend this result in presence of salt by including all cells
within a radius $\kappa^{-1}$. Then, the obtained expression is
expanded in the same two limits. The full derivation will be
detailed in a forthcoming, longer paper \cite{future_paper}.

For the limit $q \gg 1$ we find that the leading term of the
negative electrostatic persistence length of our loop expansion
and the correlation model is the same, $l_\e \simeq - q^2 l_{\rm
OSF}$.  In contrast, the fluctuation model underestimates the
electrostatic persistence length,
 yielding $l_\e \simeq l_{\rm OSF} /q^2$.

For the other limit $q = 1+ \Delta q$,  $\Delta q
\ll 1$  we find agreement between our model and the thermal
fluctuation model \cite{Golestanian}: $ l_\e \simeq  l_{\rm OSF}
[1-O(\Delta q)] $ while the correlation model overestimates it: $
l_\e \simeq l_{\rm OSF} [-1-O(\Delta q)]$. Our expansion
qualitatively accounts for both $q$ limits. There are, however,
differences in the prefactors.

Furthermore, we note that $q$, and not just the temperature $T$,
is the relevant  parameter which determines the relative
contributions of correlations compared to fluctuations. The
limits of very low or high temperatures ($\kappa \rightarrow
\infty ~ {\rm or} ~ 0$, respectively) are beyond the validity
range of our model which explicitly assumes that $L^{-1}\ll \kappa
 \ll a^{-1}$. Nevertheless, because the leading term of Eq.~(\ref{el_pl}) does not
depend on $\lb$, the electrostatic persistence length $l_{\rm e}$
neither vanishes nor diverges in these two limits. This shows
that the loop expansion method used indeed takes into account both
correlations and thermal fluctuations.

Having found the partition function, other  thermodynamically
averaged quantities can also be obtained. For instance, the
density of the bound counterions on a {\em bent} chain, or
alternatively, the effective $q$.  The effective $q$ ($q_{\rm
eff}$) is proportional to the total charge density on the PE chain
(monomers and average concentration of bound counterions). To
first order in $\psi^{\rm b}$, we find the Manning result:
$q_{\rm eff}=1$ for all chain conformations. To second order in
$\psi^{\rm b}$ the effective charge density on the chain is
smaller
\begin{equation}
   q_{\rm eff} = 1 + (q-1) \frac{1}{8 \kappa^2 \ln \kappa a} \frac{1}{\rho^2} < 1.
\label{qeff}
\end{equation}
where $\rho$ is the average cylinder radius of curvature.
On a slightly bent rod, the average distance
between condensed counterions is smaller and
correlations become more significant as compared to the straight-rod case.
This makes condensation of counterions more favorable \cite{Shklovskii}.
As the
polymer bends, more counterions are able to condense, which
in turn
 drives further reduction of the persistence length.
Note that the derivation of Eq.~(\ref{qeff})
assumes the bending of the chain
is only a small perturbation to the straight-rod limit ($ \rho \gg L$).
On a significantly bent chain ($\rho \sim L$),
the situation is completely different, and lies beyond the scope of our model.
In the limit of a folded chain resembling a spherical colloidal particle,
mean-field theory predicts that counterions are unable
to condense at all \cite{Oosawa}.

In order to
examine the effect of the increased condensation, we look at the
asymptotic form of Eq.~(\ref{el_pl}) for $q = 1+ \Delta q, \Delta
q \ll 1$ in two cases. In the first we allow the density of the
bound counterions to be adjusted according to the equilibrium
condition with the bulk (this is an expansion of
Eq.~(\ref{el_pl}) in $\Delta q$). In the second case we add a
constraint that fixes the density to be according to Manning for
all conformations of the polymer. Expanding in $\Delta q$ we
recalculate $l_\e$ for both cases
\begin{eqnarray}
    l_\e &=& l_{\rm OSF} \left[
       1 + {\rm O} ( \Delta q^2 )
        \right]
\nonumber  \\
   l_\e^{\rm fixed} &=& l_{\rm OSF} \left[
       1 - [1/ \ln (\kappa a)] \Delta q + {\rm O} ( \Delta q^2 )
        \right].
\end{eqnarray}
The persistence length is highly sensitive to the density of the multivalent
counterions. Comparison between the above two expressions
shows that corrections to Manning condensation for bent polymer
chains have a substantial influence on the persistence length.

In conclusion, we have derived an expression for the electrostatic
persistence length of a single, stiff, strongly charged PE, taking
into account both correlations and thermal fluctuations of the
bound counterions. Correlations dominate for $q \gg1$, whereas
thermal fluctuations for $q \gtrsim 1$. These two mechanisms were
considered separately in Ref.~\cite{Shklovskii} and
Ref.~\cite{Golestanian}, respectively. The advantage of our loop
expansion method is that it takes both mechanisms into account.
As a result, it covers a large range of counterion valencies and
polymer charge densities and offers a possible explanation for
the discrepancies between the two models.

Our results show a considerable
decrease in the polymer stiffness (via its persistence length)
for systems with multivalent counterions, $z\ge 2$. This decrease
depends on the modified Manning-Oosawa parameter $q$. The
estimates obtained for the collapse of semi-flexible PEs in
presence of multivalent counterions are reasonable and are
related, at least qualitatively, to the phenomenon of DNA
condensation.  The effective Hamiltonian of Eq.~(\ref{heff2}) has
the form of a three-body interaction. In the current model this
is the main source of attraction and can explain the high
sensitivity of the chain rigidity to the counterion valency $z$.

\acknowledgments
{ We wish to thank I. Borukhov, Y. Burak, H.
Diamant, A. Grosberg, B.-Y. Ha, R. Mints, R. Netz, T. Odijk, H.
Orland, B. Shklovskii and D. Thirumalai for useful discussions and
correspondence. Partial support from the U.S.-Israel Binational
Foundation (B.S.F.) under grant No. 98-00429 and the Israel
Science Foundation founded by the Israel Academy of Sciences and
Humanities
--- centers of Excellence Program is gratefully
acknowledged.}


\newpage

\begin{figure}
\onefigure[scale=0.7]{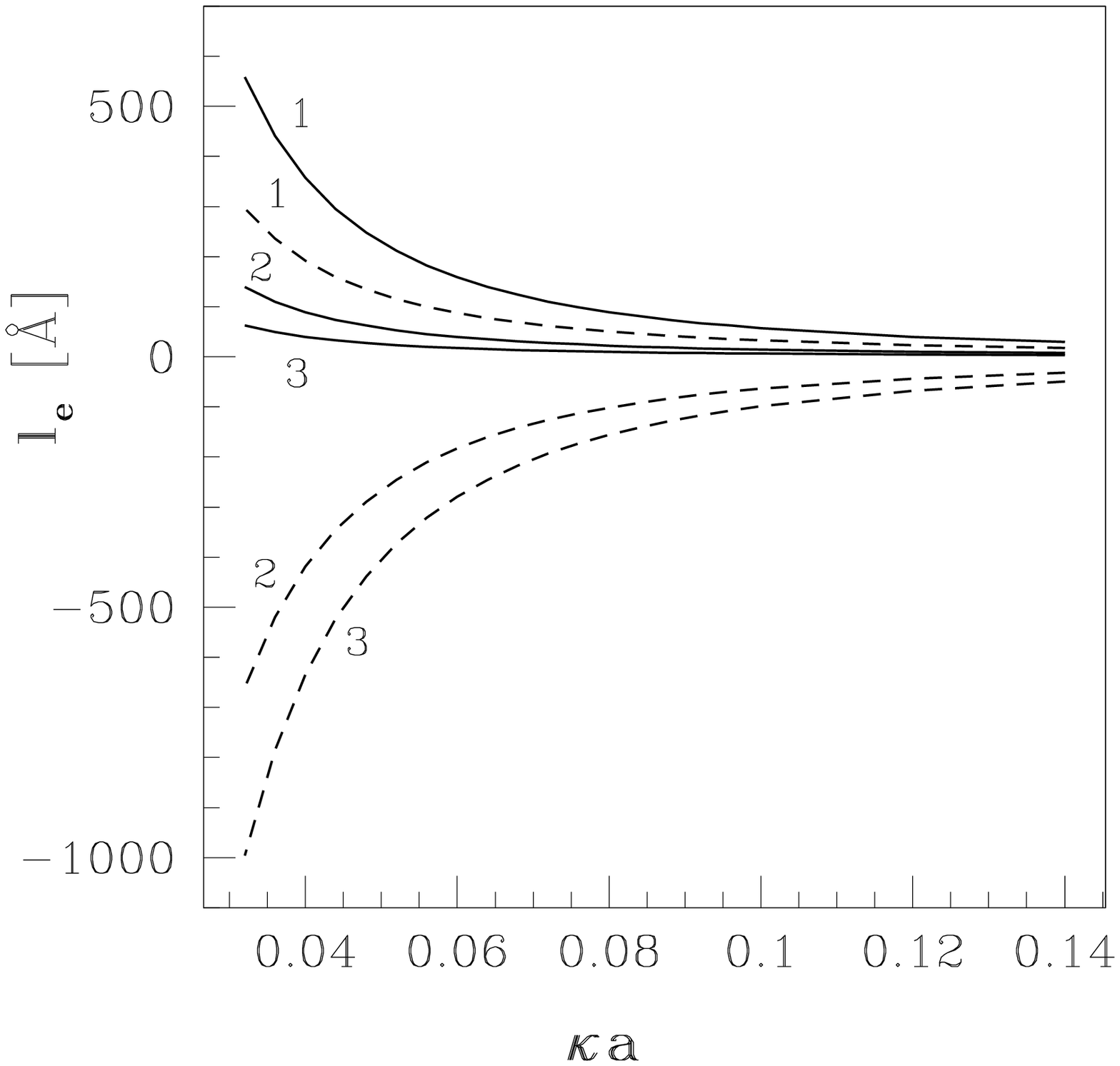} \caption{The electrostatic
persistence length $l_{\e}$ as function of $\kappa a$ according to
OSF (solid line) and our Eq.~(\ref{el_pl}) (dashed line).
Valencies are specified next to each curve. The parameters chosen
are: $a=4$\,\AA, $\lb=7$\,\AA, so that $q=1.75 z$. The negative
$l_{\rm e}$ values for $z=2,3$ indicate a possible collapse
transition of the PE chain.} \label{fig1}
\end{figure}

\end{document}